\documentstyle[prl,aps,multicol,psfig]{revtex}

\newcommand{\DMRG}{{\sc dmrg}\ }    

\newcommand{\BEQ}{\begin{equation}} 
\newcommand{\BEA}{\begin{eqnarray}}
\newcommand{\EEQ}{\end{equation}}   
\newcommand{\EEA}{\end{eqnarray}}
\newcommand{\D}{{\rm d}}            
\newcommand{\II}{{\rm i}}           
\newcommand{\rar}{\rightarrow}      
\newcommand{\wit}[1]{{\widetilde{#1}}}   
\newcommand{\ket}[1]{{\left| #1 \right>}} 

\newcommand{\matz}[4]
     {\mbox{${\begin{array}{cc} #1 & #2  \\ #3 & #4 \end{array}}$}}

\begin{document}
\draft
\title{Universal finite-size scaling amplitudes in anisotropic scaling}

\author{Malte Henkel$^a$ and Ulrich Schollw\"ock$^b$}

\address{$^a$Laboratoire de Physique des Mat\'eriaux,$^*$ Universit\'e Henri 
Poincar\'e Nancy I, \\ 
B. P. 239, F--54506 Vand{\oe}uvre-les-Nancy Cedex, France}
\address{$^b$Sektion Physik, Ludwig - Maximilians -Universit\"at M\"unchen,
Theresienstr. 37/III, D--80333 M\"unchen, Germany}

\maketitle

\begin{abstract}
Phenomenological scaling arguments suggest the existence of universal 
amplitudes in the finite-size scaling of certain correlation lengths in 
strongly anisotropic or dynamical phase transitions. For equilibrium systems,
provided that translation invariance and hyperscaling are valid,
the Privman-Fisher scaling form of isotropic equilibrium
phase transitions is readily generalized. For non-equilibrium systems, 
universality is shown analytically for directed percolation and 
is tested numerically in the annihilation-coagulation model and in the
pair contact process with diffusion. In these models, for both periodic and 
free boundary conditions, the universality of the finite-size scaling amplitude
of the leading relaxation time is checked. Amplitude universality reveals 
strong transient effects along the active-inactive transition line in the pair
contact process. 
\end{abstract}



\section{Introduction}

The notions of scaling and universality are central to the modern understanding
of critical phenomena, see e.g. \cite{Fish83,Card96,Henk99}. 
Besides the well-known universality
of the critical exponents, universality is also manifest for many critical
amplitudes, as reviewed in \cite{Priv91}. Here we are concerned with universal
amplitudes which arise in finite-size scaling. For example, 
consider a statistical system at equilibrium, e.g. a simple ferromagnet, 
which is described by an isotropic
and translation-invariant field theory in its continuum limit. Close to its
critical point, and on a lattice with finite extent $L$, Privman and Fisher
\cite{Priv84} showed that in any dimension $d$ below the upper critical
dimension $d^*$ the singular part of the free energy density $f$ and the
inverse correlation lengths $\xi_i$ satisfy the scaling form
\BEQ \label{gl:PF}
f(t,h) = L^{-d} Y (z_1, z_2) \;\; , \;\; \xi_{i}^{-1} = L^{-1} S_{i}(z_1, z_2),
\EEQ 
where $z_1 = C_1 t L^{1/\nu}$, $z_2=C_2 h L^{(\beta+\gamma)/\nu}$ are the
scaling variables, the reduced temperature $t=(T-T_c)/T_c$ and the reduced
magnetic field $h=H/T_c$ describe the distance from the critical point and
$\beta,\gamma,\nu$ are the standard equilibrium critical exponents. The
index $i$ distinguishes different correlation lengths, e.g. $i=\sigma$ for
the spin-spin correlation length or $i=\varepsilon$ for the energy-energy
correlation length in a simple ferromagnet. Furthermore,
the scaling functions $Y$ and $S_i$ are {\em universal\/} functions and all 
non-universal properties of a given model are condensed in the values of the 
non-universal metric factors $C_1$ and $C_2$. Although the functions $Y,S_i$
are universal, they do depend on the boundary conditions and of the geometric
shape of the lattices under study. 

It follows from (\ref{gl:PF}) that if the model parameters are
tuned to $t=h=0$ so that the model is at its bulk critical point, the
finite-size scaling of free energy density $f=L^{-d} Y(0,0)$ and of the 
inverse correlation lengths 
$\xi_{i}^{-1} = L^{-1} S_i(0,0)$ is described in terms of 
{\em universal\/} finite-size scaling amplitudes. This qualitative statement
can be made quantitative in $2D$ for systems defined on an infinitely long
strip of finite width $L$ through a by now classic conformal invariance
argument which for example explicitly relates the $S_{i}(0,0)$ to known
critical exponents \cite{Card84}. Concrete model studies have confirmed this
many times, as reviewed in \cite{Henk99,Priv91}. 
More recently, similar relations have
been conjectured also in $3D$ from the results of numerical studies
in toruslike \cite{Henk87,Weig99} 
and spherical geometries \cite{Card85,Weig00}.

When considering the situation of {\em dynamical\/} scaling, where
time and space scale differently, the extra degree of freedom might appear to
exclude the presence of universal finite-size scaling ampltitudes in the sense
of eq.~(\ref{gl:PF}). Here, we shall ask under what conditions 
the arguments of Privman and Fisher \cite{Priv84} 
can be generalised also to anisotropic scaling. Specifically, for a system
undergoing a phase transition with anisotropic scaling which is defined on a
lattice with finite extent $L$ in the spatial direction but is infinite in the 
temporal direction. Then (under conditions to be detailed in sections 2 and 3) 
the {\em spatial\/} correlation lengths should satisfy the scaling form
\BEQ \label{gl:GRAL}
\xi_{i,\perp}^{-1} = L^{-1} S_{i} 
\left( C_1 t L^{1/\nu_{\perp}}, C_2 h L^{(\beta+\gamma)/\nu_{\perp}}\right) 
\EEQ
in a notation analogous to (\ref{gl:PF}) and where again the $S_i$ are 
universal functions.

While the generalization to {\em equilibrium} anisotropic scaling is
a rather straightforward extension of the methods valid for the isotropic 
case \cite{Priv84}, the non-equilibrium situation is more difficult. 
We shall show how correlation length amplitude universality can be established 
for systems in the directed percolation universality class. Furthermore, 
numerical data from some reaction-diffusion models are also in agreement with 
this and suggest that the scaling form (\ref{gl:GRAL}) 
might be generally valid.

This paper is organised as follows: in section 2, we present the scaling
arguments leading to the recognition of universal amplitudes for
anisotropic scaling in equilibrium. In section 3 
non-equilibrium dynamical scaling is discussed, with emphasis on the
directed percolation universality class. In section 4,
we test the amplitude universality by analysing the finite-size scaling
of the leading inverse relaxation time in the annihilation-coagulation model
and the pair contact process with single 
particle diffusion. In section 5 we give our conclusions. An appendix contains 
the Bethe ansatz calculation of the relaxation time in an analytically 
solvable special case.

\section{Universal finite-size amplitudes in equilibrium} 

Consider an equilibrium system with anisotropic scaling in two distinct 
directions. Systems of this kind are known since a long time, e.g. in
Lifshitz points \cite{Horn75} or anisotropic uniaxial magnets \cite{Boeh79},
see \cite{Selk88,Yeom88,Selk92,Neub98} for reviews or else in
quantum phase transitions \cite{Hert76}, see \cite{Sach99} for a recent book.
One can derive the phenomenological scaling of the physical observables from 
the covariance of the correlators under scale transformations and 
reconstruct the thermodynamics this way, see \cite{Card96,Henk99}. 
In line with conventional phenomenological treatments of 
strongly anisotropic scaling \cite{Bind89,Leun91}, we assume for 
the two-point functions the scaling form
\BEQ \label{gl:ZP}
G_{i}\left( r_{\perp},r_{\|};t,h\right) 
= b^{-2x_{i}} 
G_{i}\left( \frac{r_{\perp}}{b}, \frac{r_{\|}}{b^{\theta}}; 
t b^{y_t}, h b^{y_h}
\right)
\EEQ
where $b$ is the rescaling factor, $t,h$ refer to physical quantities like the
reduced temperature and the reduced magnetic field, $\theta$ is the
{\em anisotropy} exponent and $y_t, y_h, x_i$ are scaling exponents. The index
$i$ refers to different physical quantities of which the two-point function
is formed, e.g. $i=\sigma$ for the spin operator or $i=\varepsilon$ for the 
energy density (for simplicity, we use throughout a notation analogous to 
simple ferromagnets). At criticality, $t=h=0$, and one has
\BEA
G_{\sigma}(r_{\perp},0) \sim r_{\perp}^{-2x_{\sigma}}\; &,&
\; G_{\sigma}(0,r_{\|}) \sim r_{\|}^{-2x_{\sigma}/\theta} ,
\nonumber \\
G_{\varepsilon}(r_{\perp},0) \sim r_{\perp}^{-2x_{\varepsilon}}\; &,&
\; G_{\varepsilon}(0,r_{\|}) \sim r_{\|}^{-2x_{\varepsilon}/\theta} . 
\EEA
For a strongly anisotropic
equilibrium system, $r_{\perp}$ and $r_{\|}$ correspond to different
directions in space. This case is for example realised at the Lifshitz point
in spin systems with competing interactions 
like the ANNNI model \cite{Selk92,Dieh00}.  
For brevity, we shall refer to the directions $r_{\perp}$ as 
{\em `spatial'} and to the directions $r_{\|}$ as {\em `temporal'}. 

We use the equilibrium fluctuation-dissipation theorem 
\BEQ \label{gl:FlucDiss}
\chi = \int\! \D r_{\|} \D^d r_{\perp}\: G_{\sigma}(r_{\perp}, r_{\|}) 
\;\; , \;\;
C = \int\! \D r_{\|} \D^d r_{\perp}\: 
G_{\varepsilon}(r_{\perp}, r_{\|}) ,
\EEQ
where $d$ is the number of `spatial' dimensions and $\chi,C$ are the 
susceptibility and specific heat. We shall work with a
single `temporal' direction throughout, but generalisations are obvious. Units
are such that the critical temperature $T_{\rm c}=1$. From (\ref{gl:ZP})
and integrating, one gets immediately the scaling forms for 
$\chi=\chi(t,h)=-\partial^2 f/\partial h^2$
and for $C=C(t,h)=-\partial^2 f/\partial t^2$. Here $f$ is the (singular part)
free energy density. From the scaling of $\chi$ and $C$, it should
satisfy the scaling
\BEQ
f(t,h) = b^{\theta+d-2x_{\sigma}-2y_{h}} f(t b^{y_t}, h b^{y_h}) 
= b^{\theta+d-2x_{\varepsilon}-2y_{t}} f(t b^{y_t}, h b^{y_h}) .
\EEQ
These two forms are consistent, if $x_{\sigma}+y_{h}=x_{\varepsilon}+y_{t}=w$. 
In fact, the above argument works for any pair of scaling operators
like $\sigma,\varepsilon$ and their conjugated scaling fields $h,t$. Therefore,
the value $w$ must be independent of all physical scaling operators which
might be present in a given model. Next, we define the standard static critical
exponents $\alpha,\beta,\gamma$ as usual, e.g. \cite{Card96}. 
Also, out of criticality, we expect an exponential 
decrease of the two-point function, characterised
by the correlation lengths $\xi_{\perp,\|} \sim t^{-\nu_{\perp,\|}}$ 
(at $h=0)$. From this, 
we find that $\nu_{\perp}=1/y_{t}$, $\theta=\nu_{\|}/\nu_{\perp}$ and
$\beta+\gamma=y_{h}/y_{t}$. From dimensional counting, we expect $w=d+\theta$
and the free energy density then scales as
\BEQ
f(t,h) = b^{-d-\theta} f(t b^{y_t}, h b^{y_h}) .
\EEQ
As usual, we explicitly assume the absence of dangerously 
irrelevant scaling fields \cite{Fish83,Priv84,Bind89}. We then recover 
the {\em hyperscaling} relation $2-\alpha=\nu_{\|}+d\nu_{\perp}$. 
The common scaling form for the thermodynamics is found by scaling 
out $b$ and introducing the conventional critical exponents:
\BEQ \label{gl:freiE}
f(t,h) = A_1 |t|^{2-\alpha} W^{\pm}\left( A_2 h |t|^{-\beta-\gamma} \right) ,
\EEQ
where $W^{\pm}$ are the universal scaling functions which are obtained for
$t>0$ and $t<0$, respectively, and $A_{1,2}$ are non-universal 
metric constants. At this level, the anisotropy of the scaling of the
two-point function only appears in the generalized form of the hyperscaling
relation. 

Equation (\ref{gl:freiE})  will be the starting point 
for our discussion of finite-size scaling. In what follows, 
we consider a situation where the `spatial' directions
are of finite extent $L$ whereas the `temporal' direction remains infinite. 
In the same spirit as Privman and Fisher \cite{Priv84}, we assume that the
{\em finite-size} scaling behaviour is governed by the `spatial' correlation
length $\xi_{\perp}$ only and we write
\BEQ \label{gl:freiFSS}
f(t,h;L) = A_1 |t|^{2-\alpha} 
W^{\pm}\left( A_2 h |t|^{-\beta-\gamma}; L\xi_{\perp}^{-1} \right),
\EEQ
where the {\em bulk} `spatial' correlation length 
$\xi_{\perp} =\xi_0 t^{-\nu_{\perp}}$. Note that there is {\em no} extra
metric factor in the second argument of $W^{\pm}$, whereas $\xi_0$ is
non-universal. To simplify the notation, we assume that there is no phase
transition for $L$ finite, but this restriction could be removed analogously
to the equilibrium case \cite{Sing87}. We emphasize that the `temporal' 
direction remains infinite, otherwise we would have to deal with two
{\em distinct\/} finite length scales. 

Following the ideas developed by Privman and Fisher \cite{Priv84}, we have to
trace the non-universal constants, taking into account the
anisotropic scaling. For that, it is sufficient
to study the `spatially' infinite system. We expect the scaling form of the
connected spin-spin correlator, see also (\ref{gl:ZP}),
\BEQ
G_{\sigma}(r_{\perp},r_{\|};t,h) = D_0 D_1 r_{\perp}^{-2x_{\sigma}}
X^{\pm}\left( r_{\perp}/\xi_{\perp},D_0 r_{\|}/\xi_{\perp}^{\theta};
D_2 h |t|^{-\beta-\gamma}\right) ,
\EEQ
where $X^{\pm}$ is a universal scaling function and $D_{0,1,2}$ are 
non-universal metric factors. From the fluctuation-dissipation theorem 
(\ref{gl:FlucDiss}) one has
\BEQ \label{gl:chiEins}
\chi(t,h) = D_1 \xi_{\perp}^{\gamma/\nu_{\perp}} \wit{X}^{\pm}
\left(D_2 h |t|^{-\beta-\gamma}\right) .
\EEQ
where $\wit{X}^{\pm}$ is a new scaling function obtained from $X^{\pm}$. 
Now, for the {\em non-connected} spin-spin correlator, one has in the same way,
introducing new universal scaling functions $Z^{\pm}$
\BEQ
\Gamma_{\sigma}(r_{\perp},r_{\|};t,h) = D_0 D_1 r_{\perp}^{-2x_{\sigma}}
Z^{\pm}\left( r_{\perp}/\xi_{\perp},D_0 r_{\|}/\xi_{\perp}^{\theta};
D_2 h |t|^{-\beta-\gamma}\right)
\EEQ
At this point, we assume translation invariance with respect 
to  $r_{\perp}$ and $r_{\|}$. 
Therefore, there should exist a mean magnetization $m$ which is independent
of $r_{\perp}$ and $r_{\|}$ and which can be found
considering $\Gamma_{\sigma}$ at large separations $r_{\perp},r_{\|}$: 
\BEA
\Gamma_{\sigma} (r_{\perp}, r_{\|}; t,h) &=& 
D_0 D_1 \xi_{\perp}^{-2x_{\sigma}} 
Z^{\pm}\left( 1, D_0 r_{\|}/r_{\perp}^{\theta};
D_2 h |t|^{-\beta-\gamma}\right)  \\
\downarrow~~~~~~ & & ~~~~~~~~~~~\downarrow \nonumber \\
m^2(t,h) &=& D_0 D_1 \xi_{\perp}^{-2x_{\sigma}} \wit{Z}^{\pm}\left( 
D_2 h |t|^{-\beta-\gamma}\right), \label{gl:magEins}
\EEA
where the arrow indicates taking the limit of 
large `spatio-temporal' separations. Because of translation invariance, 
$m^2$ should become independent of $D_0$. 
On the other hand, applying standard thermodynamics to the free energy
(\ref{gl:freiE}) yields
\BEA
m(t,h) &=& A_1 A_2 |t|^{\beta} 
           W_1^{\pm}\left( A_2 h |t|^{-\beta-\gamma}\right), 
\label{gl:magZwei}\\
\chi(t,h) &=& A_1 A_2^2 |t|^{-\gamma} 
              W_2^{\pm}\left( A_2 h |t|^{-\beta-\gamma}\right),
\label{gl:chiZwei}
\EEA
where $W_{n}^{\pm}(x) = \D^n W^{\pm}(x)/\D x^n$. 

We now compare eqs.~(\ref{gl:chiEins}) and (\ref{gl:chiZwei}). Letting
first $h=0$, we find
\BEQ
D_1 \xi_{0}^{\gamma/\nu_{\perp}} = A_1 A_2^2 U_1 .
\EEQ
Comparing the arguments of the scaling functions, we have
\BEQ
D_2 = A_2 U_2 .
\EEQ
Next, we compare eqs.~(\ref{gl:magEins}) and (\ref{gl:magZwei}) and find for
$h=0$ that
\BEQ
D_0 D_1 \xi_{0}^{-2\beta/\nu_{\perp}} = A_1^2 A_2^2 U_3
\EEQ
(since $x_{\sigma}=\beta/\nu_{\perp}$). Here, $U_{1,2,3}$ are {\em universal}
constants whose universality follows from the universality of the scaling
functions considered. Using the hyperscaling relation
$\gamma+2\beta=(d+\theta)\nu_{\perp}$, we find that
\BEA
A_1 \xi_{0}^{d+\theta} D_0^{-1} &=& Q_1 = U_1/ U_3 \nonumber \\
D_2 A_2^{-1}           &=& Q_2 = U_2 \label{gl:Qconst} \\
D_0^{\gamma/(\nu_{\perp}(d+\theta))} 
D_1 A_1^{-1-\gamma/(\nu_{\perp}(d+\theta))} A_2^{-2} &=& Q_3 = 
U_1^{1-\gamma/(\nu_{\perp}(d+\theta))} U_3^{\gamma/(\nu_{\perp}(d+\theta))}
\nonumber 
\EEA
and the $Q_{1,2,3}$ are universal constants. 

Finally, we come back to the finite-size scaling behaviour. 
In eq.~(\ref{gl:freiFSS}), we replace $\xi_0$ by $A_1$ using (\ref{gl:Qconst}).
Scaling out $L$ and using again hyperscaling, it is easy to arrive at the
scaling form
\BEQ \label{gl:Yskal}
f(t,h;L) = L^{-d-\theta} D_0
Y\left( C_1 t L^{1/\nu_{\perp}}, C_2 h L^{(\beta+\gamma)/\nu_{\perp}}\right),
\EEQ
where $Y$ is a universal scaling function and $C_{1,2}$ are
non-universal metric factors related to $A_{1,2}$. In contrast with the
isotropic situation (\ref{gl:PF}), we see that the finite-size scaling
amplitude of the free energy is no longer universal. Furthermore, since
$\xi_{\|} = \xi_{\perp}^{\theta}/D_0$, we have 
\BEQ \label{gl:fxipxipp}
f(t,0;L) \xi_{\perp}^{d}(t,0;L)\xi_{\|}(t,0;L) = 
D_0^{-1} f(t,0;L) \xi_{\perp}^{d+\theta}(t,0;L) \mathop{\rar}_{t\to0} 
\mbox{\rm univ. constant},
\EEQ
which holds because of (\ref{gl:Qconst}). Therefore, we expect for the
{\em `spatial'} correlation length
\BEQ \label{gl:Sskal}
\xi_{\perp}^{-1} = L^{-1} 
S\left( C_1 t L^{1/\nu_{\perp}}, C_2 h L^{(\beta+\gamma)/\nu_{\perp}}\right)
\EEQ
with a {\em universal} scaling function $S$ and the {\em same} metric factors
$C_{1,2}$ as in (\ref{gl:Yskal}). While this analysis was phrased in terms
of the transverse spin-spin correlation 
length $\xi_{\perp}=\xi_{\perp,\sigma}$, similar arguments
should hold for the `spatial' correlation lengths $\xi_{\perp,i}$ of any other
physical observable, with $S$ in (\ref{gl:Sskal}) being replaced by an
appropriate function $S_i$. The scaling functions $Y$ and $S_i$ should depend
on the boundary conditions and, if $d\geq 2$, on the shape of the finite
`spatial' domain. Note that $L$ refers here to the physical length, which cannot
be equated to the number of sites $N$ times the lattice constant in non-square
lattices \cite{Priv84,Sing87}. For a recent example of this in the $2D$ 
Ising model context, see \cite{Izma00}. 

Finally, the `temporal'
correlation lengths $\xi_{\|,i}$ should read 
\BEQ \label{gl:Temp}
\xi_{\|,i}^{-1} = L^{-\theta} D_0 
R_i\left( C_1 t L^{1/\nu_{\perp}}, C_2 h L^{(\beta+\gamma)/\nu_{\perp}}\right)
\EEQ
with universal scaling functions $R_i$ and again the {\em same} metric factors
$C_{1,2}$ as before. 
The value of $D_0$ is related to the dimensionful constant which occurs
in the energy-momentum dispersion relation of the underlying continuum field
theory and cannot be found straightforwardly. 
However, at criticality ($t=h=0$) ratios of `temporal'
correlation lengths $\xi_{\|,i}/\xi_{\|,j}$ should tend to universal
constants in the $L\to\infty$ limit. The universality of these ratios would 
not have immediately been obvious from straightforward finite-size scaling. 

Eq.~(\ref{gl:Sskal}) is the main result of this section. It
provides the natural generalisation of the Privman-Fisher form (\ref{gl:PF})
to the case of equilibrium anisotropic scaling. It is immediate to
include further physical parameters into the analysis. 
We emphasize that (i) translation invariance and 
(ii) hyperscaling was required in deriving 
this result. We stress that we considered finite sizes in the 
`spatial' direction and obtain universality for the spatial correlation
lengths $\xi_{\perp,i}$ only. 

\section{Universal finite-size amplitudes out of equilibrium}

We have seen that in equilibrium systems with anisotropic scaling and in a
geometry where the `spatial' directions are of finite extent $L$ while the
`temporal' directionis infinite, the
`spatial' correlation lengths have a universal finite-size scaling amplitude.
We now ask whether this result generalizes towards more general forms of
dynamical scaling, without appealing to the special properties of
equilibrium systems. Fluctuations in non-equilibrium systems can be treated
in terms of dynamic functionals via Martin-Siggia-Rose theory 
\cite{Mart73,Baus76,Domi76,Domi78}. To be specific, we shall work in a setting 
of reaction-diffusion processes, of which directed percolation is a common
example, see \cite{Schm95,Marr99,Hinr99,Matt98} for reviews. 
We shall continue to denote time by $r_{\|}$ and space by
$r_{\perp}$. As before, $t$ measures the distance from the steady-state
critical point and $h$ denotes an external field (e.g., for directed
percolation $t=p-p_c$ and $h$ is the rate of a process $\emptyset\to A$). 
For the sake of technical simplicity, we shall assume translation invariance
throughout. As in equilibrium, we have to trace the non-universal metric
factors and this is most conveniently done in the bulk. 

Physical quantities of interest are the mean particle 
density $\rho$, the survival probability $P$ and the pair connectedness 
function $G=G(r_{\perp}',r_{\|}';r_{\perp},r_{\|})$, which is defined as the
probability that the sites $(r_{\perp}',r_{\|}')$ and $(r_{\perp},r_{\|})$ are
connected by a direct path \cite{Marr99,Hinr99}. 
Because of translation invariance
$G=G(r_{\perp}'-r_{\perp},r_{\|}'-r_{\|})$, which will be used throughout. 
These quantities are expected to satisfy the scaling behaviour
\BEA
\rho(r_{\perp},r_{\|};t,h) &=& b^{-x_{\rho}} \rho\left( 
\frac{r_{\perp}}{b},\frac{r_{\|}}{b^z}; t b^{y_t}, h b^{y_h}\right) 
=D_{1\rho}\,\xi_{\perp}^{-x_{\rho}} 
{\cal E}^{\pm}\left( \frac{r_{\perp}}{\xi_{\perp}},
D_0 \frac{r_{\|}}{\xi_{\perp}^z}; D_2 h |t|^{-y_h/y_t}\right) \nonumber\\
P(r_{\perp},r_{\|};t,h) &=& b^{-x_{P}} P\left( 
\frac{r_{\perp}}{b},\frac{r_{\|}}{b^z}; t b^{y_t}, h b^{y_h}\right) 
=D_{1P}\, \xi_{\perp}^{-x_{P}} 
{\cal F}^{\pm}\left( \frac{r_{\perp}}{\xi_{\perp}},
D_0 \frac{r_{\|}}{\xi_{\perp}^z}; D_2 h |t|^{-y_h/y_t}\right) \label{gl:rPG}\\
G(r_{\perp},r_{\|};t,h) &=& b^{-x_{G}} G\left( 
\frac{r_{\perp}}{b},\frac{r_{\|}}{b^z}; t b^{y_t}, h b^{y_h}\right) 
=D_{1G}\, \xi_{\perp}^{-x_{G}} {\cal G}^{\pm}
\left( \frac{r_{\perp}}{\xi_{\perp}},
D_0 \frac{r_{\|}}{\xi_{\perp}^z}; D_2 h |t|^{-y_h/y_t}\right) \nonumber
\EEA
where the $x$'s are scaling dimensions and $y_{t,h}$ renormalization group
eigenvalues, the $D$'s are 
non-universal metric factors, ${\cal E},{\cal F},{\cal G}$ are universal
scaling functions where the index distinguishes between the cases $t>0$ and
$t<0$, $\xi_{\perp}=\xi_0 |t|^{-\nu_{\perp}}$ is the spatial, 
$\xi_{\|}=\xi_{\perp}^z /D_0$ is the temporal correlation length and $z$ is
the dynamical exponent (as before $y_t=1/\nu_{\perp}$). 

In the steady state, and for $h=0$, one expects $\rho\sim t^{\beta}$ and
$P\sim t^{\beta'}$. In general, the two exponents $\beta$ and $\beta'$ are 
distinct from each other. For spatial translation invariance, the dependence
on $r_{\perp}$ drops out for both $\rho$ and $P$ and in the steady state
(i.e. $r_{\|}\to\infty$) one has
\BEA
\rho(t,h) &=& D_{1\rho}\,\xi_{0}^{-\beta/\nu_{\perp}} 
\wit{\cal E}^{\pm}\left(  D_2 h |t|^{-y_h/y_t}\right) |t|^{\beta} \nonumber\\
P(t,h) &=& D_{1P}\,\xi_{0}^{-\beta'/\nu_{\perp}} 
\wit{\cal F}^{\pm}\left(  D_2 h |t|^{-y_h/y_t}\right) |t|^{\beta'}
\label{gl:DichteWkeit} 
\EEA
where $x_{\rho}=\beta/\nu_{\perp}$, $x_{P}=\beta'/\nu_{\perp}$ 
and $\wit{\cal E}^{\pm}=\lim_{r_{\|}\to\infty} {\cal E}^{\pm}$ 
and similarly for $\cal F$. We also consider the auto-connectedness (that is 
$r_{\perp}=r_{\perp}'$) in the steady state
\BEQ 
G(0,\infty;t,h) =: G(t,h) = D_{1P}\,\xi_{0}^{-x_G} 
\wit{\cal G}^{\pm}\left(  D_2 h |t|^{-y_h/y_t}\right) |t|^{x_G \nu_{\perp}}
\EEQ
In the active phase ($t>0$), the surviving clusters will create an average
density $\sim |t|^{\beta}$ in the interior of the spreading cone. Therefore,
the auto-connectedness should in the steady state saturate at the value
\cite{Mend94}
\BEQ \label{gl:GrP}
G(t,h) = \rho(t,h) P(t,h)
\EEQ 
Comparison of the scaling forms then yields, setting $h=0$,
\BEQ \label{gl:GrhoP}
x_G = (\beta+\beta')/\nu_{\perp} \;\; , \;\;
D_{1G} = D_{1\rho}D_{1P}\, 
\frac{\wit{\cal E}^{\pm}(0) \wit{\cal F}^{\pm}(0)}{\wit{\cal G}^{\pm}(0)}
\EEQ
Usually, $x_G=d-\theta z$ is expressed in terms of the initial critical
slip exponent $\theta$ \cite{Jans89}, 
which makes it apparent that the expression (\ref{gl:GrhoP}) is
in fact a generalized hyperscaling relation \cite{Marr99,Hinr99,Mend94}. 

Next, we consider the total mass $M$ of the cluster \cite{Hinr99}, given by
\BEQ\label{gl:Masse}
M(t,h) := \int_{\rm I\!R^d} \!\D^d r_{\perp} \int_{0}^{\infty} \!\D r_{\|}\,
G(r_{\perp},r_{\|};t,h) = \frac{D_{1G}}{D_0} \xi_{\perp}^{\gamma/\nu_{\perp}} 
\overline{\cal G}^{\pm}\left( D_2 h |t|^{-y_h/y_t}\right)
\EEQ
where eq.~(\ref{gl:rPG}) was used and $\overline{\cal G}^{\pm}$ is a new 
universal function related to ${\cal G}^{\pm}$. Also
\BEQ \label{gl:hypgamma}
\gamma = d\nu_{\perp} + \nu_{\|} - \beta -\beta' 
\EEQ
which is the analogue of the hyperscaling relation of the equilibrium systems. 

While the discussion so far has been completely general, we now appeal to
two properties which are valid for systems in the directed percolation
universality class, but need not be generically valid. First, we consider
a directed percolation process in the presence of a weak field $h$
(physically, $h$ parametrises the rate of a particle creation process
$\emptyset\to A$). A site
at a given time becomes active if it is connected with at least one active site
in the past, where a particle was created by the field. The number of such
sites is equal to the cluster size, the probability to become active
is given by the density \cite{Hinr99}
\BEQ
\rho(t,h) \simeq 1 - (1-h)^{M(t,h)} \simeq h M(t,h)
\EEQ
for $h$ small. Therefore, 
\BEQ \label{gl:Msuszep}
M(t,0) = \left. \frac{\partial \rho(t,h)}{\partial h}\right|_{h=0}
\EEQ
Comparison with the scaling forms for $\rho$ and $M$ leads to
\BEQ \label{gl:DDD}
y_h/y_t = \beta+\gamma \;\; , \;\;
D_{1P} = D_0 D_2 \xi_{0}^{-(\beta+\gamma)/\nu_{\perp}} {\cal A}^{\pm}
\EEQ
where ${\cal A}^{\pm}$ is an universal amplitude. Second, directed percolation
is special in the sense that there is a `duality' symmetry which 
can be used to show that \cite{Gras79}
\BEQ \label{gl:dual}
\rho(t,h) = P(t,h)
\EEQ
As a consequence, $\beta=\beta'$ and $D_{1\rho}=D_{1P}$ for directed 
percolation and we thus have, combining 
eqs.~(\ref{gl:DichteWkeit},\ref{gl:Masse},\ref{gl:DDD})
\BEA
\rho(t,h) &=& D_0 D_2 \xi_{0}^{-d-z} |t|^{\beta} 
\hat{\cal M}_1^{\pm}\left( D_2 h |t|^{-\beta-\gamma}\right)
\nonumber \\
M(t,h) &=& D_0 D_2^2 \xi_{0}^{-d-z} |t|^{-\gamma} 
\hat{\cal M}_2^{\pm}\left( D_2 h |t|^{-\beta-\gamma}\right)
\EEA
with universal functions $\hat{\cal M}_n^{\pm}(x)
=\D^n \hat{\cal M}^{\pm}(x)/\D x^n$ and where the hyperscaling relation 
eq.~(\ref{gl:hypgamma}) has been used. We therefore recover the analogues
of eqs.~(\ref{gl:magZwei},\ref{gl:chiZwei}) found in equilibrium.   
Finally, we define a new function $\mu=\mu(t,h)$ by 
$\rho(t,h) = \partial \mu(t,h)/\partial h$, which implies
\BEQ
\mu(t,h) = D_0 \xi_{0}^{-d-z} |t|^{(d+z)\nu_{\perp}} \hat{\cal M}^{\pm}
\left( D_2 h |t|^{-\beta-\gamma}\right)
\EEQ
In particular, as we did before at equilibrium, we have because of
$\xi_{\|}=\xi_{\perp}^z/D_0$ that
\BEQ \label{gl:univmu}
\mu(t,0) \xi_{\perp}^d(t,0) \xi_{\|}(t,0) \mathop{\rar}_{t\to0} 
\mbox{\rm univ. constant},
\EEQ
which is indeed the analogue of the result (\ref{gl:fxipxipp}).

At last, we consider a geometry of finite size $L$ in space but of infinite
extent in time. Again, by analogy with section 2, 
we postulate that in this finite geometry merely 
the scaling functions are modified
\BEQ
\hat{\cal M}_n^{\pm} = \hat{\cal M}_n^{\pm} 
\left( D_2 h |t|^{-\beta-\gamma}; L\xi_{\perp}^{-1}\right) 
\EEQ
and without introducing any further metric factor. 
Indeed, we can then scale out $L$ and, because of eq.~(\ref{gl:univmu}),
arrive at the same scaling forms
(\ref{gl:Sskal},\ref{gl:Temp}) as had been found before for the spatial
and temporal correlation lengths in anisotropic
equilibrium systems, at least for systems in the directed percolation
universality class. Since directed percolation is known to be equivalent to
an equilibrium (in fact, purely geometrical) problem, the existence of
universal finite-size amplitudes in this class is not too surprising
and might have been anticipated from the discussion in the previous section. 

While the spatial correlation length $\xi_{\perp}$ may not be always a very
accessible quantity, its universality may also be tested by considering the
spatial moment (let $t=h=0$ for simplicity)
\BEA
R_L^{(n)} := \langle r_{\perp}^n\rangle
&=& \frac{\int_{\Lambda(L)}\!\D^d r_{\perp}\int_{0}^{\infty}\!\D r_{\|}\, 
r_{\perp}^n G(r_{\perp},r_{\|};L/\xi_{\perp})}
{\int_{\Lambda(L)}\!\D^d r_{\perp}\int_{0}^{\infty}\!\D r_{\|}\, 
G(r_{\perp},r_{\|};L/\xi_{\perp})} \nonumber\\
&=& \xi_{\perp}^n
\frac{\int_{\Lambda(L/\xi_{\perp})}
\!\D^d r_{\perp}\int_{0}^{\infty}\!\D r_{\|}\,
r_{\perp}^{n-x_G} {\cal G}^{\pm}(r_{\perp},r_{\|};L/\xi_{\perp})}
{\int_{\Lambda(L/\xi_{\perp})}\!\D^d r_{\perp}\int_{0}^{\infty}\!\D r_{\|}\,
r_{\perp}^{-x_G} {\cal G}^{\pm}(r_{\perp},r_{\|};L/\xi_{\perp})}
\nonumber \\
&=& \xi_{\perp}^n \Xi_{n}(L/\xi_{\perp})\nonumber \\
&=& L^n \wit{\Xi}_{n}(L/\xi_{\perp})
\EEA
where $\Lambda(L)$ is a $d$-dimensional hypercube of linear extent $L$ and 
${\Xi}_n$ and $\wit{\Xi}_n$ are universal functions. 
Since there is no metric factor in 
the argument of $\wit{\Xi}_n$, the universality of the finite-size scaling
amplitude of $\xi_{\perp}$ is equivalent to the universality of the finite-size
scaling amplitude of $R_L^{(n)}$ (on the other hand, the temporal moment
$\langle r_{\|}^n\rangle \sim (D_0^{-1} L^z)^n$ has a non-universal amplitude).
The universality of these moments is a somewhat stronger statement than the
universality of certain ratios of moments $\langle \rho^n\rangle$ of the
particle density $\rho$ which has
recently been verified in $1D$ and in $2D$ for several models in the directed 
percolation universality class \cite{Dick98}. 

In summary, we have seen that for systems in the directed percolation 
universality class, the special properties 
eqs.~(\ref{gl:Msuszep},\ref{gl:dual}), taken together with the general 
relation (\ref{gl:GrP}), are sufficent to rederive the universal 
finite-size scaling form (\ref{gl:Sskal}) of the spatial correlation length,
in spite of the absence of the fluctuation-dissipation relation. It is not
yet clear wether there exist more general arguments which would permit us to
arrive at the same result without appealing to either (\ref{gl:Msuszep}) or 
(\ref{gl:dual}). However, we shall in the next section present numerical 
evidence that the universal finite-size scaling forms (\ref{gl:Sskal}) for
$\xi_{\perp}$ or (\ref{gl:Temp}) for $\xi_{\|}$ might be more generally valid. 

\section{Reaction-diffusion processes}

The new information contained in (\ref{gl:Sskal}) which goes beyond the
standard RG ideas is the universality of
the finite-size scaling amplitude $L \xi_{\perp}^{-1}$ precisely at 
criticality (and similarly the universality of {\em all} 
ratios $\xi_{\|,i}/\xi_{\|,j}$).
Since for the time being, the derivation of this universality for 
non-equilibrium systems appears to be restricted to directed percolation, 
we use the pair contact process and the
annihilation-coagulation model to test the universality 
hypothesis advanced in sections 2 and 3 quantitatively.

The pair contact process \cite{Jens93} 
has been intensively studied recently. It is a
reaction-diffusion system, where particles move and react on a lattice. While
each lattice site can be either empty or be occupied by a single particle, the
following microscopic moves are permitted
\begin{eqnarray} 
\left\{
\begin{array}{ccc}
AA \emptyset & \rightarrow & AAA \\
\emptyset AA & \rightarrow & AAA
\end{array}
\right.
\,\,\,\,\,\,\, &{\rm with \,\,\, rate}& 
\,\,\,\,\,\,\,  \frac{(1-p)(1-d)} 2
\nonumber \\
AA~ \rightarrow  ~\emptyset \emptyset~~~~~~~~~~ 
  &{\rm with \,\,\,rate}& \,\,\,\,\,  p(1-d)  
\label{gl:PCPD}\\
A \emptyset~ \leftrightarrow  ~\emptyset A~~~~~~~~~~
   &{\rm with \,\,\,rate}& \,\,\,\,\,  d  \nonumber
\end{eqnarray} 
and are parametrized by the diffusion constant $d$ and the pair annihilation 
rate $p$. 

\begin{figure}[h]
\centerline{\psfig{file=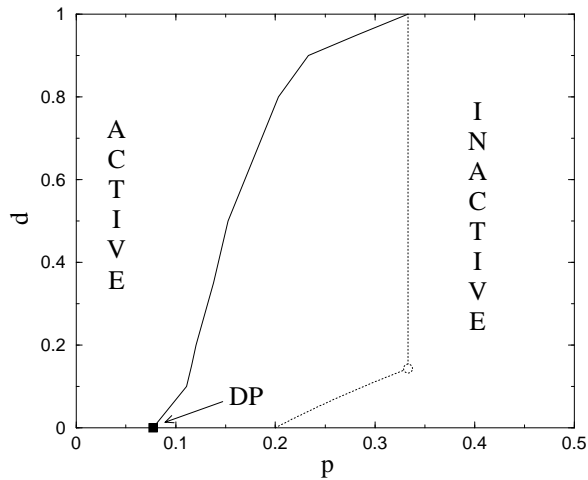,height=6.5cm}}
\vskip 0.2truecm
\caption{Steady state phase diagram of the pair contact process. 
The dotted lines are the phase boundaries according to 
pair mean field theory, while the full line gives the active-inactive
transition in $1D$. DP marks the steady-state transition in the directed
percolation universality class at $d=0$.}
\label{phasediagram}
\end{figure}

While in the case without diffusion ($d=0$), 
the steady-state transition between the
active and the absorbing state was found to be in the directed percolation
universality class \cite{Jens93,Dick98,Kamp99}, 
the effects of adding diffusion were first
studied using field-theoretical methods, considering a bosonic
field theory without any restriction on the number of particles per 
site (which leads to a divergent particle density in the 
active phase)\cite{Howa97}. It was shown that the entire absorbing phase
is critical and in the universality class of diffusion-annihilation 
(see below). Because
of the non-renormalizability of the underlying field theory, no quantitative
information about the transition towards the active state could be obtained.
The first quantitative informations were obtained \cite{Carl00} 
through the use of density matrix
renormalization group ({\sc dmrg}) techniques \cite{Whit92,Pesc99,Carl99}.
The steady state phase diagram is shown in figure~\ref{phasediagram} 
and there is a general
agreement between \DMRG and Monte Carlo studies 
on the location of the critical point $p_c(d)$
\cite{Carl00,Mend99,Hinr00,Gras00,Odor00}. 

The annihilation-coagulation model is formulated in the same way, with the
allowed reactions
\begin{eqnarray} 
AA~ \rightarrow  ~A \emptyset~,~\emptyset A ~~~ 
  &{\rm with \,\,\,rate}& \,\,\,\,\,  d \gamma  \nonumber \\
AA~ \rightarrow  ~\emptyset \emptyset~~~~~~~~~~~ 
  &{\rm with \,\,\,rate}& \,\,\,\,\,  2 d \alpha \label{gl:anco} \\
A \emptyset~ \leftrightarrow  ~~\emptyset A~~~~~~~~~~
   &{\rm with \,\,\,rate}& \,\,\,\,\,  d \nonumber 
\end{eqnarray} 
parametrized by $\alpha$ and $\gamma$ for annihilation and coagulation,
respectively. The long-time behaviour is the model is always
algebraic, i.e. the mean particle density $\rho(t)\sim t^{-1/2}$ in $1D$, see
\cite{Schm95,Marr99,Hinr99,Matt98,Schu00} and references therein. 

In this paper, we shall use the Hamiltonian formulation of reaction-diffusion
processes \cite{Kada68,Doi76,Gras80,Peli85,Alca94}, which starts from 
the master equation 
\begin{equation} 
\frac{\partial | P(t) \rangle}{\partial t} = - H | P(t) \rangle
\end{equation} 
where $| P(t) \rangle$ a state vector and $H$ is referred to as
``quantum'' Hamiltonian (for recent reviews, see \cite{Hinr99,Schu00}). 
For a chain with $L$ sites, 
$H$ is a stochastic $2^L \times 2^L$ matrix with elements
\begin{equation}
\langle \sigma |H |\tau \rangle = - w(\tau\to\sigma) \;\; , \;\;
\langle \sigma |H |\sigma\rangle = \sum_{\tau\neq\sigma} w(\sigma\to\tau)
\end{equation}
where $|\sigma\rangle$, $|\tau\rangle$ are the state vectors of the particle
configurations $\sigma,\tau$ and $w$ are the transition rates. It is well-known
that the ground state energy of the pair contact process 
$E_0 = E_1 = 0$ is twofold degenerate for $d\ne 0$ \cite{Carl00}.
The energy gap $\Gamma = E_2 - E_0$, calculated from the second excited 
state, is the inverse relaxation time or, in the notation of
section 2, the inverse `temporal' correlation length $\xi_{\|}^{-1}=\Gamma$ 
towards the steady state.  
We shall consider both free and periodic boundary conditions. 

{\bf 1.} First, we discuss the pair contact process (\ref{gl:PCPD}). 
We find the following, surprisingly simple, finite-size scaling behaviour
for the gap $\Gamma_L$ {\em in the entire absorbing phase}, that is, for all
$p\geq p_c(d)$, namely
\BEQ \label{eq:ConjG}
\Gamma_L = a d L^{-2} \left(1 + O\left( L^{-1}\right) \right)
\EEQ
where $a$ depends on the boundary condition
\BEQ \label{eq:Conja}
a = \left\{ \matz{2\pi^2}{\mbox{\rm ~~~~~~;~ periodic}}
{\pi^2}{\mbox{\rm ;~ free}} \right.
\EEQ
but is {\em independent} of both $p$ and $d$.

Before deriving (\ref{eq:Conja}), we argue that this result
confirms the universality of the correlation length
amplitudes discussed in section 3. Indeed, for systems in the
diffusion-annihilation universality class already the inverse `temporal' 
correlation length $\Gamma$ has a universal amplitude, provided that 
the value of the diffusion constant $d$ is fixed. The universality of the
reaction-diffusion process $2A\to\emptyset$ has been discussed using
field theory methods \cite{Peli86,Droz93,Lee94}. From the 
renormalisation group equations, it can be shown
that the value of the diffusion constant $d$ is not renormalised through the
effects of the interaction of the particles and simply stays at its bare value.
The bare value of $d$ is the value it has in the original lattice formulation 
of the problem. Since the diffusion constant sets the time scale, we expect for 
the gap $\Gamma \sim L^{-2} d$. These calculations \cite{Peli86,Droz93,Lee94} 
apply to the process $2A\to\emptyset$ which 
corresponds to the case $p=1$ in the 
model at hand. However, it is known that in the entire absorbing phase,  
the extra interactions coming from the reaction $2A\to 3A$ are 
irrelevant \cite{Howa97}. Therefore, they should not modify the value of $d$
(since we only consider here the inactive phase, we leave aside the question
how $d$ evolves under renormalization in the active phase). 
Consequently, the proportionality of $\Gamma$ and $d$ in eq.~(\ref{eq:ConjG}) 
comes from the non-renormalisation of $d$. That non-renormalisation is
a special property of the diffusion-annihilation universality class. 
That is consistent with the scaling form (\ref{gl:Temp}) for $\xi_{\|}$ and
we can identify $D_0=d$. 
Finally, the $p$-independence of the amplitude $a$ in eq.~(\ref{eq:Conja}) 
is an example of the universality of the finite-size scaling amplitude
$R(0,0)$ in eq.~(\ref{gl:Temp}), which in turn implies the universality of
$S(0,0)$ in eq.~(\ref{gl:Sskal}).

We now derive eqs.~(\ref{eq:ConjG},\ref{eq:Conja}). 
We need the lowest
non-vanishing eigenvalue $\Gamma=E_2$ of the quantum Hamiltonian $H$. For 
$p=1$, it turns out that the spectrum of $H$ is equal to the spectrum of an
XXZ Heisenberg quantum chain $H_{\rm XXZ}$. 
The lowest gap of $H_{\rm XXZ}$ can be
found from the coordinate Bethe ansatz \cite{Alca87,Alca88}. This reproduces
eq.~(\ref{eq:Conja}) for $p=1$ and all values of $d$. The details of the 
calculation are presented in the appendix. 

We point out, however, that
for $d=1/2$ and free boundary conditions only, the Bethe ansatz equations
have a closed-form solution. The {\em exact\/} lowest gap for any finite 
number of sites $L$ is
\BEQ \label{eq:Deltahalb}
\Gamma_L = 1 - \cos \frac{\pi}{L+1} \;\; ; \;\; d = \frac{1}{2} \;\; , \;\; 
p = 1 \;\; , \;\; \mbox{\rm free b.c.}
\EEQ
in agreement with (\ref{eq:ConjG},\ref{eq:Conja}). 
This had been conjectured before on the basis of numerical data 
\cite{Carl99} (closed-form solutions for slightly different $\Delta=1/2$ 
XXZ chains have been discussed recently in \cite{Frid00}).

For $p\ne 1$, the PCPD is not related to any known integrable model and we
revert to numerical methods. We consider the normalized amplitudes
\BEQ
A_L^{({\rm P})} = L^2 \Gamma_L/(2\pi^2 d) \;\; , \;\; 
A_L^{({\rm F})} = L^2 \Gamma_L/(\pi^2 d)
\EEQ
defined for periodic (P) and free (F) boundary conditions, respectively. If
and only if (\ref{eq:Conja}) is correct, the amplitudes $A_L^{({\rm P})}$ and 
$A_L^{({\rm F})}$ should converge towards unity in the $L\to\infty$ limit. 

In table~\ref{tabper} we show data for periodic boundary conditions. These
were obtained from diagonalizing $H$ through the standard Arnoldi 
algorithm \cite{Henk99}. Translation and parity invariance were used to 
blockdiagonalize $H$, and lead to 
matrices of size $\approx 2^L/(2L)$ for $L$ sites.
In table~\ref{tabfrei} data for free boundary conditions are shown. They were
obtained by applying the density matrix renormalization group ({\sc dmrg})
method \cite{Whit92,Pesc99,Carl99} to the pair contact process \cite{Carl00}.

Clearly, the data for both $A_L^{({\rm P})}$ and $A_L^{({\rm F})}$ at finite
values of $L$ are in general quite far away from unity. We also see that the
raw data tend to be closer to unity for smaller values of $d$. However, since
the systematic variation of these amplitudes with $L$ is huge, a precise
$L\to\infty$ extrapolation must be performed. We have used the 
BST extrapolation algorithm \cite{Buli64} which has established itself as a
reliable and precise method for the extrapolation of finite-lattice sequences
arising in both equilibrium and non-equilibrium critical phenomena. The 
parameter $\omega$ describes the (effective) leading finite-size correction of 
a sequence $A_L = A_{\infty} + {\cal A}_1 L^{-\omega}+\ldots$ and must be 
chosen to optimize convergence \cite{Henk99,Buli64}. 

In all cases, we find that the extrapolated amplitudes
\BEQ
A_{\infty}^{({\rm P})} := \lim_{L\to\infty} A_L^{({\rm P})} \simeq 1 \;\; , \;\;
A_{\infty}^{({\rm F})} := \lim_{L\to\infty} A_L^{({\rm F})} \simeq 1 
\EEQ
within the numerical accuracy of the extrapolation and in full agreement
with eq.~(\ref{eq:Conja}). The need for $L\to\infty$ extrapolation also means 
that the universal infinite-size amplitudes may be hard to see in, say, 
Monte Carlo simulations. We also give the effective value of $\omega$ for each
sequence. In view of the exact result $\omega=1$ for $p=1$, 
see (\ref{gl:kP},\ref{gl:kF}), it is satisfying that $\omega$ stays 
close to one. 

In general, for a given $d$ convergence is best for $p$ close to 1 and decreases
when $p$ is lowered. The lowest values of $p$ given in table~\ref{tabper}
are more or less the smallest ones for which a reliable convergence of the
amplitudes could still be observed. In varying $d$, we see that data converge
best for relatively small $d$ and that close to $d=1$ the cross-over towards 
mean-field behaviour \cite{Carl00} affects the finite-size scaling of the 
amplitudes. Comparing the data for periodic and free boundary conditions for the
same values of $p$ and $d$, we observe that the $A_{\infty}^{({\rm F})}$ are 
closer to unity than the $A_{\infty}^{({\rm P})}$ for the same value of $L$
(and in contrast to the usual expectation that finite-size corrections should
be smallest for periodic boundary conditions). For $p=1$, these remarks are
confirmed analytically, see eqs.~(\ref{gl:kP},\ref{gl:kF}). 

All in all, the extrapolated amplitudes converge
over a large range of values of $p$ and $d$ towards unity, in agreement with 
(\ref{eq:Conja}). Therefore, the PCPD in the inactive phase confirms the
universality of the correlation length finite-size scaling amplitude as
derived in section 2. Since the entire inactive phase is expected to be
in the same universality class, that result should
apply even to those portions of the inactive phase where our relatively
short chains did not permit us to carry out a precise extrapolation.

{\bf 2.} Second, we briefly discuss the 
annihilation-coagulation model (\ref{gl:anco}). 
It is well known \cite{Peli86,Kreb95a,Henk95,Simo95,Balb95} 
that the quantum Hamiltonian 
$H=H(\alpha,\gamma)$ is similar to the quantum Hamiltonian $H(\alpha+\gamma,0)$
of pure annihilation. Therefore eqs.~(\ref{eq:ConjG},\ref{eq:Conja}) 
also apply to this model, independently of $\alpha$ and $\gamma$.
If we take $\alpha+\gamma=1$, the steady-state
particle density amplitude 
$\lim_{L\to\infty}L\rho(L) = (1+\alpha/\gamma)/(1+2\alpha/\gamma)$ 
depends on the branching ratio $\alpha/\gamma$ and is not universal. 

Eqs.~(\ref{eq:ConjG},\ref{eq:Conja}) state that for  
the finite-size amplitude
of $\Gamma$ is independent of the ratio $r=\alpha+\gamma$ 
of the reaction rate and the diffusion rate $d$. 
In the light of the universality hypothesis of sections 2 and 3, the observed 
$r$-independence of the amplitude $L^2 \Gamma$ 
means that the critical exponents of the pair 
annihilation process $2A\to\emptyset$ (or the equivalent coagulation
process $2A\to A$ \cite{Peli86,Kreb95a}) should also be independent of $r$, 
i.e. the mean particle density $\rho(t)\sim t^{-\delta}$ with $\delta=1/2$. 
While that had been anticipated long ago by many people, exact lattice
calculations only exist for $r=1$, see \cite{Hinr99,Schu00}. The only published 
verifications of the $r$-independence of $\delta$ we are aware of either used 
purely numerical methods \cite{Simo95,Kreb95,Gryn98},
a real-space renormalisation group scheme \cite{Hooy00} or other {\it ad hoc}
approximations \cite{Bare99}.  
 
The steady-state particle density should scale as
\BEQ
\rho_L =  D_0 C_2 L^{-\beta/\nu_{\perp}} 
\left. Y'\left( 0, C_2 h L^{d+\theta-\beta/\nu_{\perp}}\right)\right|_{h=0}
\EEQ
(borrowing a notation from section 2, 
where the prime indicates the derivative with respect to the second argument
and $h$ parametrises a source of particles). That is standard finite-size
scaling without any readily identifiable universal amplitude and is therefore
weaker than the forms of sections 2 and 3. It is known from field theory
that $C_2$ is independent of $r$ \cite{Droz93} while it does depend on the 
branching ratio $\alpha/\gamma$ \cite{Balb95} which is compatible with our
results.

{\bf 3.} Having checked the scaling function universality 
in some examples with
known behaviour, we now illustrate how this universality might be
used as a diagnostic tool. Appealing to the experience gathered in 
equilibrium systems \cite{Priv91}, universal amplitudes might be expected to
vary considerably more between distinct universality classes than critical
exponents. Therefore, even an approximate determination of universal
amplitude may allow to conclude on the universality class of the model at hand. 
Reconsider the phase diagram of the pair
contact process in fig.~\ref{phasediagram}. Presently, there is no consensus
on how many universality classes should be realised along the transition line
between the active and inactive phases for $d\ne 0,1$. As a starting point, 
we might consider pair mean field theory, which predicts two distinct 
universality classes along the two segments of the pair mean field 
transition curve \cite{Carl00}. 
The calculation of both steady-state and time-dependent
critical exponents in $1D$ from simulations \cite{Odor00} appear to be in
agreement with this prediction. On the other hand, {\sc dmrg} studies 
\cite{Carl00} and different simulations in $1D$ \cite{Hinr00,Gras00,Hinr00a}
only find evidence for a single universality class along the transition line. 

\begin{figure}[h]
\centerline{\psfig{file=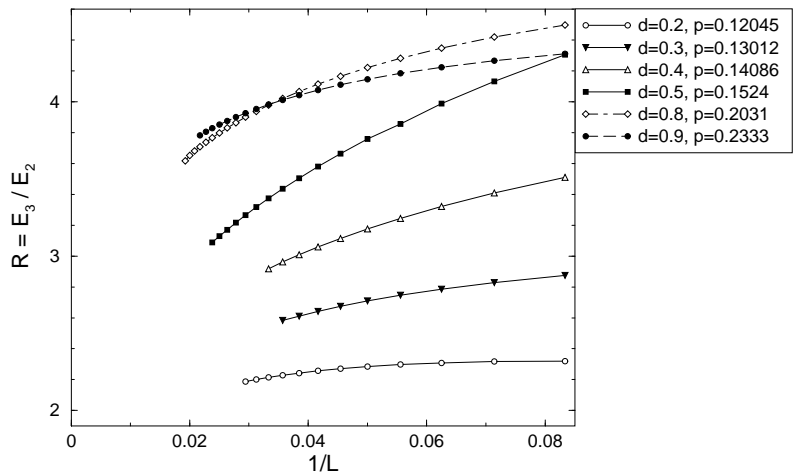,height=6.5cm}}
\vskip 0.2truecm
\caption{Ratio $R=E_3/E_2$ of the two lowest eigenvalues of $H$ at the critical
line $p=p_c(d)$ of the pair contact process, for free boundary conditions and
several values of $d$.}
\label{amplitude}
\end{figure}

Here, we consider the ratio $R = E_3 / E_2$ of the two lowest non-vanishing 
eigenvalues of $H$ for free boundary conditions as obtained
from the {\sc dmrg}. We use the Monte Carlo estimates for $p_c(d)$ obtained
by Grassberger \cite{Gras00}. 
$R$ is equal to the ratio of two distinct relaxation times 
and from eq. (\ref{gl:Temp}), we expect $R$ to be constant within a given 
universality class. Our numerical results are shown in fig.~\ref{amplitude}.

At first sight, it might be possible to separate the 
values of $R$ into two classes, one 
for smaller values of $d$ (up to $d\approx 0.4 -0.5$) and a limit 
$R_{\infty}\sim 2$, and one for larger values of $d$ (above $d\approx 0.8$)
with a limit $R_{\infty}\sim 3-4$. The fact that $R_{\infty}$ is quite 
independent of $d$ for $0.2\leq d\leq 0.4$ confirms the expected universality
($R_{\infty}= 1$ in the inactive phase because there all levels are two-fold
degenerate, see appendix).

However, closer inspection reveals that already for $d=0.5$, 
the data for $R_L$ for
$L$ small start out close to 4 and then begin to cross over to values close to
2. That signals the presence of strong transient effects in this model. 
In addition, for the values of $L$ for which data exist, the values
of $R$ for both $d=0.8$ and $d=0.9$ are close together and quite close to 4. 
However, these data, in particular for $d=0.8$, also show indications that 
they might also cross over to 
smaller values of $R$ if $L$ could be increased further. The lattice sizes
available are not large enough to be able to distinguish clearly between the 
possibilities of a single \cite{Carl00,Hinr00,Gras00,Hinr00a} 
or two \cite{Odor00} transitions, although the possibility of a single 
transition appears more likely. If there is a change of the universality
class along the critical line at all, fig.~\ref{amplitude} suggests that it 
certainly should occur for $d > 0.5$. 
 
\section{Conclusions}

In this paper, we have tried to generalize Privman-Fisher \cite{Priv84} 
universality to (steady-state) phase transitions with dynamical scaling. 
For equilibrium systems the standard arguments carry over, the main ingredients
being translation invariance and hyperscaling. With respect to more standard
renormalisation group arguments, only the absence of non-universal metric
factors in front of the scaling function for $\xi_{\perp,i}$ 
in eq.~(\ref{gl:Sskal}) is new. Equivalently, this may be stated, see  
eq.~(\ref{gl:Temp}), as the universality of {\em all} 
ratios $\xi_{\|,i}/\xi_{\|,j}$. 

Out of equilibrium, new arguments must be sought. In the special
case of directed percolation, the peculiar properties
eqs.~(\ref{gl:Msuszep},\ref{gl:dual}) were seen to be sufficient for 
amplitude universality. These properties may or may not be available in other
universality classes, but we found some numerical evidence in several 
reaction-diffusion systems that the universal finite-size scaling 
forms (\ref{gl:Sskal},\ref{gl:Temp}) might indeed hold in general. In the 
annihilation-coagulation model and the pair contact process (inactive phase), 
we found that the finite-size scaling amplitude of the leading relaxation time 
is indeed independent of the irrelevant parameters we considered. Further 
evidence in favour of universality was found by studying the active-inactive
transition line in the pair contact process. In addition, our data for the 
ratio of the two leading relaxation times appear to favour a single
universality class along that transition line for $0<d<1$. It remains an
open question how to derive the universal scaling forms 
(\ref{gl:Sskal},\ref{gl:Temp}) in general non-equilibrium systems.  

Recently, the reaction-diffusion process $2A\to A$ and $A\emptyset A\to 3A$
has been studied \cite{Henk00}. 
If the coagulation rate is equal to the diffusion rate, the
model is exactly solvable. It remains in the universality class of 
diffusion-annihilation for all values of the particle production rate $\lambda$.
{}For periodic boundary conditions, the exact amplitude of the  
leading inverse relaxation time is 
$\lim_{L\to\infty} L^2 \Gamma_L = 2\pi^2 d$, independently 
of $\lambda$, as to be expected from (\ref{eq:ConjG},\ref{eq:Conja}).

Finally, upon identification the universal finite-size scaling
amplitude of the transverse `spatial' correlation lengths $\xi_{\perp,i}$, 
it might be tempting to ask if, in analogy to 
equilibrium \cite{Card84,Henk87,Weig99,Card85,Weig00}, there
could be relations of universal finite-size scaling amplitudes with some 
exponents. To answer this question would require a set of worked-out examples 
on which some hypothesis of this kind could be tried out.

\acknowledgements
We thank E. Carlon for his contributions in developing the application of the
\DMRG method to non-equilibrium systems, P. Grassberger for kindly
sharing his data on $p_c(d)$ before publication, H. Hinrichsen for discussions
and L. Turban for a critical reading of the manuscript.  
MH thanks the Complexo Interdisciplinar of the University of Lisbon for
warm hospitality. 
This work was supported by the French-German Procope programme. 

\section*{Appendix: Bethe ansatz calculation}

We calculate the lowest gap $\Gamma=E_2$ for $p=1$ in the pair contact process 
and derive the amplitude $a$ in (\ref{eq:Conja}). 

In the $p=1$ case,
only the pair annihilation $AA\rar \emptyset\emptyset$ survives. It is well
known \cite{Alca94} that in this case, the quantum Hamiltonian can be 
decomposed $H = H_0 + H_1$ in such a way that the eigenvalue spectrum of $H$ is 
independent of $H_1$, viz. ${\rm spec\ }(H) = {\rm spec\ }(H_0)$. The latter
is related to the spectrum of the XXZ Heisenberg chain which 
can be found from the coordinate Bethe ansatz \cite{Alca87,Alca88}. 

We begin with the {\em periodic} case. The spectrum-generating part of $H$ is
\BEQ
H_0 = d H_{\rm XXZ}(\Delta,t) + \frac{1}{4}(1+d)L
\EEQ
where
\BEQ
H_{\rm XXZ}(\Delta,t) = -\frac{1}{2} \sum_{i=1}^{L} \left[
\sigma_{i}^x\sigma_{i+1}^x+\sigma_{i}^y\sigma_{i+1}^y+
\Delta\sigma_{i}^z\sigma_{i+1}^z+t\sigma_{i}^z \right]
\EEQ
and
\BEQ \label{eq:DeltaT}
\Delta = \frac{3d-1}{2d} \;\; , \;\; t=\frac{1-d}{d}
\EEQ
Since the total spin $S^z := \sum_{i=1}^{L} \sigma_i^z$, commutes with
$H_{\rm XXZ}$, the eigenstates $H_0\ket{\psi_n}=E_n\ket{\psi_n}$ 
can be classified in terms of the number $n$ of reversed spins, viz. 
$S^z \ket{\psi_n} = (L-2n)\ket{\psi_n}$. The lowest 
states with $n=0,1$ have zero
energy and correspond to the two steady states of the model. The lowest gap
$\Gamma$ is found in the sector $n=2$. From \cite{Alca88}, one has
\BEQ
E_2 = 2d( 2 - \cos k -\cos k')  \;\; ,  \;\; 
L k = 2\pi I - \Theta(k,k') \;\; , \;\; L k' =2\pi I' - \Theta(k',k)
\EEQ
where
\BEQ \label{eq:ThetaFunk}
\Theta(k,k') = 2\arctan \frac{\Delta\sin((k-k')/2)}
{\cos((k+k')/2) - \Delta\cos((k-k')/2)}
\EEQ
and $I,I'=\pm\frac{1}{2},\pm\frac{3}{2},\ldots$ 
are distinct half-integers. The total momentum of the corresponding
state is $P=k+k'=2\pi(I+I')/L'$. The lowest energy gap will have 
$P=0$, or $k'=-k$. Furthermore, the lowest energy state corresponds to
the choice $I=-I'=1/2$, as can be checked by considering the special
case $\Delta=0$ (because of the symmetry between $k$ and $k'$, all levels
are two-fold degenerate). We find
\BEQ \label{eq:LuePer}
\Gamma = 4d( 1 -\cos k ) \;\; , \;\; 
\tan \frac{Lk-\pi}{2} = -\frac{\Delta\sin k}{1-\Delta \cos k}
\EEQ
For $L$ large, the solution of the second equation (\ref{eq:LuePer}) is
\BEQ \label{gl:kP}
k \simeq \frac{\pi}{L} \left( 1 - \frac{2\Delta}{1-\Delta}\frac{1}{L} + \ldots
\right)
\EEQ
Inserting this into (\ref{eq:LuePer}), we arrive indeed at the first case 
of eqs.~(\ref{eq:ConjG},\ref{eq:Conja}). 

Second, we consider {\em free} boundary conditions. The spectrum-generating
part of $H$ is
\BEQ
H_0 = d H_{\rm XXZ}(\Delta,t,r) + \frac{1}{4}(1+d)(L-1)
\EEQ
\BEQ
H_{\rm XXZ}(\Delta,t,r) = -\frac{1}{2} \left\{
\sum_{i=1}^{L-1} \left[ \sigma_{i}^x\sigma_{i+1}^x+\sigma_{i}^y\sigma_{i+1}^y+
\Delta\sigma_{i}^z\sigma_{i+1}^z \right] 
+ r\left(\sigma_1^z+\sigma_L^z\right) 
+ t \sum_{i=1}^{L} \sigma_i^z \right\}
\EEQ
where we used (\ref{eq:DeltaT}) and $r=-t/2$. Again, $\Gamma$ will be the 
lowest energy in the $n=2$ sector. From the Bethe ansatz \cite{Alca87}, one has
\BEQ \label{eq:BF}
E_2 = 2d(2-\cos k - \cos k') \;\; , \;\; 
e^{2\II(L-1)k} = \left( \frac{f_{-k}(r,\Delta)}{f_{k}(r,\Delta)}\right)^2
e^{-\II\Theta(k,k')+\II\Theta(-k,k')} 
\EEQ
where $f_k(a,b)=a-b+e^{\II k}$. A similar equation holds for $k'$, 
where $k$ and $k'$ are exchanged with respect to (\ref{eq:BF}). 
The lowest excitations in the $n=2$ sector are found for 
$k'=0$ (up to a two-fold degeneracy). 
Since $r-\Delta=-1$, using (\ref{eq:ThetaFunk}) 
and taking the logarithm, we find 
\BEQ \label{eq:LueFrei}
\Gamma = 2d(1-\cos k) \;\; , \;\; 
Lk=\pi(I+1)-2\arctan\left(\frac{\Delta}{1-\Delta}\tan\frac{k}{2} \right)
\EEQ
where $I=0,1,2,3,\ldots$. The lowest gap is obtained for $I=0$, as can be
checked for $\Delta=0$. In analogy to the periodic case, we
rewrite the second equation in (\ref{eq:LueFrei}) in the form
$\tan((Lk-\pi)/2) =-\Delta/(1-\Delta)\tan(k/2)$. 
For $L$ large, the solution is
\BEQ \label{gl:kF}
k \simeq \frac{\pi}{L} \left( 1 - \frac{\Delta}{1-\Delta}\frac{1}{L} 
+\ldots \right)
\EEQ
from which the second case in eqs.~(\ref{eq:ConjG},\ref{eq:Conja}) follows. 

Evidently, for $\Delta=0$, that is $d=1/3$, 
we recover for both boundary conditions 
the well-known results found from free-fermion methods. 
A second closed solution exists for $d=1/2$ and {\em free} boundary conditions. 
Then $\Delta=1/2$ and the second of eqs. (\ref{eq:LueFrei}) reduces to 
$(L+1)k=\pi(I+1)$. The lowest gap is obtained for $I=0$ and we arrive at
(\ref{eq:Deltahalb}). We also remark that for the $\Delta=1/2$ XXZ chain with
boundary terms such that $U_q(sl(2))$ invariance holds, the Bethe ansatz
equation can be solved analytically \cite{Frid00}.

\newpage 
\begin{table}[t]
\caption{Estimates for the normalized amplitude 
$A_L^{({\rm P})} = L^2 \Gamma_L /( 2\pi^2 d)$
for several values of $p$ and $d$ and periodic boundary conditions. The line
labeled $\infty$ gives the $L\to\infty$ extrapolations obtained from the
BST extrapolation algorithm {\protect\cite{Buli64}} and $\omega$ is the 
effective correction exponent used in these extrapolations. 
In some cases the sequences are not monotonous, 
this is indicated by a $^*$ at the value of $\omega$ used.\label{tabper}}
\begin{tabular}{|c|c|cccc|} \hline
 & & \multicolumn{4}{c|}{$p$} \\ \hline
$d=0.10$ & $L$ & 0.16 & 0.25 & 0.50 & 0.75                        \\\cline{2-6}
&    6 & 0.7274475706 & 1.1014160997 & 1.5543363873 & 1.7166144988\\
&    8 & 0.6791992625 & 0.9951704118 & 1.3575537598 & 1.4819774944\\
&   10 & 0.6559859469 & 0.9505431819 & 1.2648992305 & 1.3631372237\\
&   12 & 0.6480035923 & 0.9318542215 & 1.2110759662 & 1.2913069885\\
&   14 & 0.6483265043 & 0.9247019281 & 1.1757710267 & 1.2431938517\\
&   16 & 0.6530990879 & 0.9229800785 & 1.1507547386 & 1.2087157132\\
&   18 & 0.6602551037 & 0.9238966184 & 1.1320630670 & 1.1827963304\\
&   20 & 0.6686545657 & 0.9260886593 & 1.1175466079 & 1.1626011550\\\cline{2-6}
&$\infty$ & 1.04(3)   & 1.00(3)      & 0.9999(1)    & 1.0001(2)   \\\cline{2-6}
&$\omega$ & 1$^*$     & 1.11$^*$     &     1        & 1.0032      \\\hline
\multicolumn{6}{c}{ } \\\hline
$d=0.30$ & $L$ & 0.20 & 0.30 & 0.50 & 0.75                        \\\cline{2-6}
&    6 & 0.3393736439 & 0.5565119224 & 0.8448999662 & 1.0205743378\\
&    8 & 0.3708396385 & 0.6016754416 & 0.8664772095 & 1.0164775547\\
&   10 & 0.4001274702 & 0.6385279575 & 0.8836745934 & 1.0137328840\\
&   12 & 0.4287823558 & 0.6698979484 & 0.8975656827 & 1.0118226252\\
&   14 & 0.4562391902 & 0.6967097472 & 0.9088380138 & 1.0104117821\\
&   16 & 0.4820573931 & 0.7197719798 & 0.9180744129 & 1.0093198954\\
&   18 & 0.5060692639 & 0.7397682858 & 0.9257311017 & 1.0084452275\\
&   20 & 0.5282757026 & 0.7572450703 & 0.9321538891 & 1.0077262607\\\cline{2-6}
&$\infty$ & 0.96(4)   & 1.0001(1)    & 1.0000(3)    & 0.99999(3)\\\cline{2-6}
&$\omega$ & 1.085     & 1.074        &  1           & 1         \\\hline
\multicolumn{6}{c}{ } \\\hline
$d=0.50$ & $L$ & 0.30 & 0.40 & 0.50 & 0.99                        \\\cline{2-6}
&    6 & 0.2634463944 & 0.3594043077 & 0.4362871259 & 0.5950441871\\
&    8 & 0.3071022508 & 0.4156753699 & 0.4975263719 & 0.6582403419\\
&   10 & 0.3480300086 & 0.4649432503 & 0.5485668186 & 0.7054235500\\
&   12 & 0.3861353468 & 0.5076282318 & 0.5909715317 & 0.7416603392\\
&   14 & 0.4210763888 & 0.5445274246 & 0.6264865185 & 0.7702258735\\
&   16 & 0.4528601529 & 0.5765616531 & 0.6565615085 & 0.7932568153\\
&   18 & 0.4816890281 & 0.6045559890 & 0.6823105808 & 0.8121848167\\
&   20 & 0.5078408510 & 0.6291933618 & 0.7045790759 & 0.8279970372\\\cline{2-6}
&$\infty$ & 1.00(2)   & 1.007(2)     & 0.996(2)     & 1.0002(2)   \\\cline{2-6}
&$\omega$ & 0.98      &  1           & 1.067        & 0.994       \\\hline
\multicolumn{6}{c}{ } \\\hline
$d=0.90$ & $L$ & 0.50 & 0.75 & 0.90 & 0.95                        \\\cline{2-6} 
&    6 & 0.0541722289 & 0.0729000818 & 0.0782390633 & 0.0790103464\\
&    8 & 0.0671363692 & 0.0907802309 & 0.0975809620 & 0.0985881954\\
&   10 & 0.0801954402 & 0.1084170765 & 0.1165093810 & 0.1177191004\\
&   12 & 0.0931636953 & 0.1256311545 & 0.1348893588 & 0.1362791328\\
&   14 & 0.1059635357 & 0.1423692854 & 0.1526874700 & 0.1542393436\\
&   16 & 0.1185589168 & 0.1586177799 & 0.1699027666 & 0.1716014095\\
&   18 & 0.1309311714 & 0.1743785534 & 0.1865468998 & 0.1883788379\\
&   20 & 0.1430702131 & 0.1896605667 & 0.2026368456 & 0.2045900568\\\cline{2-6}
&$\infty$ & 1.01(1)   & 0.999(2)     & 1.001(3)     & 1.001(2)    \\\cline{2-6}
&$\omega$ & 1.081     & 1.08         & 1.078        & 1.079       \\\hline
\end{tabular}
\end{table}

\newpage
\begin{table}[t]
\caption{Estimates for the normalized amplitude 
$A_L^{({\rm F})} = L^2 \Gamma_L /(\pi^2 d)$
for several values of $p$ and $d$ and free boundary conditions, and the 
$L\to\infty$ extrapolation. When the \DMRG algorithm did not yield stable
results the corresponding finite-size entries are left empty. \label{tabfrei}}
\begin{tabular}{|c|c|ccc|} \hline
 & & \multicolumn{3}{c|}{$d$} \\ \hline
$p=0.40$ & $L$ & 0.333        & 0.50         & 0.666        \\ \cline{2-5}
         & 12  & 0.8572577379 & 0.6725747038 & 0.4583935580 \\
         & 14  & 0.8737590556 & 0.7053840790 & 0.4981815051 \\
         & 16  & 0.8869945040 & 0.7322840138 & 0.5327541136 \\
         & 18  & 0.8978251485 & 0.7547785176 & 0.5630353770 \\
         & 20  & 0.9068361253 & 0.7738292422 & 0.5897524204 \\
         & 22  & 0.9144401079 & 0.7901798864 & 0.6134826888 \\
         & 24  & 0.9209358777 & 0.8043583176 & 0.6346890106 \\
         & 26  & 0.9265447049 & 0.8167680285 & 0.6537455074 \\
         & 28  & 0.9314335748 & 0.8277177891 & 0.6709574987 \\
         & 30  & 0.9357318117 & 0.8374491423 & 0.6865761177 \\
         & 32  & 0.9395733083 & 0.8461522447 & 0.7008078882 \\
         & 34  & 0.9429388328 & 0.8539906968 & 0.7138314424 \\\cline{2-5}
     &$\infty$ & 0.999(3)     & 0.9999(5)    & 0.9995(9)    \\\cline{2-5}
     &$\omega$ & 1.08         & 1.0245       & 1.029        \\\hline
\multicolumn{5}{c}{ } \\\hline
$p=0.50$ & $L$ & 0.333        & 0.50         & 0.666        \\\cline{2-5}
         & 12  & 0.9127350184 & 0.7395843960 & 0.5254151900 \\ 
         & 14  & 0.9239698644 & 0.7686830752 & 0.5656063523 \\
         & 16  & 0.9327349799 & 0.7919832846 & 0.5996949840 \\
         & 18  & 0.9397547158 & 0.8110625710 & 0.6289496021 \\
         & 20  & 0.9454946130 & 0.8269700162 & 0.6543147959 \\
         & 22  & 0.9502701952 & 0.8404325820 & 0.6765067894 \\
         & 24  & 0.9543020404 & 0.8519708198 & 0.6960780276 \\
         & 26  & 0.9577542440 & 0.8619676543 & 0.7134610277 \\
         & 28  & 0.9607290383 & 0.8707097768 & 0.7289990177 \\
         & 30  & 0.9633226275 & 0.8784202612 & 0.7429679437 \\
         & 32  &              & 0.8852717697 & 0.7555906209 \\
         & 34  &              & 0.8913924472 & 0.7670514417 \\\cline{2-5}
    &$\infty$  & 0.9996(6)    & 0.9996(5)    & 0.9996(5)    \\\cline{2-5}
    &$\omega$  & 1.0925       & 1.0165       & 1.0191       \\\hline
\end{tabular}
\begin{tabular}{|c|c|cccc|} \hline
 & & \multicolumn{4}{c|}{$p$} \\ \hline
$d=0.90$ & $L$ & 0.60         & 0.70         & 0.80         & 0.90\\\cline{2-6}
         & 12  & 0.1863167451 & 0.2042528770 & 0.2172237514 & 0.2255705980\\  
         & 14 & 0.2119456015  & 0.2317569546 & 0.2459600505 & 0.2549873770\\  
         & 16 & 0.2362199311  & 0.2576333440 & 0.2728737999 & 0.2824706979\\  
         & 18 & 0.2592237747  & 0.2820034649 & 0.2981179611 & 0.3081867029\\  
         & 20 & 0.2810372384  & 0.3049794681 & 0.3218249399 & 0.3322848660\\  
         & 22 & 0.3017367417  & 0.3266645014 & 0.3441189613 & 0.3548989577\\  
         & 24 & 0.3213939696  & 0.3471482795 & 0.3651117354 & 0.3761534335\\  
         & 26 & 0.3400759780  & 0.3665336490 & 0.3849046790 & 0.3961552283\\ 
         & 28 & 0.3578455055  & 0.3848833645 & 0.4035899034 & 0.4150046251\\  
         & 30 & 0.3748295723  & 0.4022778186 & 0.4212504431 & 0.4327914966\\  
         & 32 & 0.3908774672  & 0.4187863074 & 0.4378639789 & 0.4495863889\\  
         & 34 & 0.4062441268  & 0.4344650723 &              &  \\\cline{2-6}
    &$\infty$ & 0.986(20)     & 0.98(3)    & 0.983(15) & 0.984(15)\\\cline{2-6}
    &$\omega$ & 1.09          & 1.103        & 1.100        & 1.099 \\\hline
\end{tabular}
\end{table}

\end{document}